\documentclass[12pt]{article}
\newcommand{\fr}{ \frac}
\newcommand{\beq}{\begin{equation}}
\newcommand{\eeq}{\end{equation}}
\newcommand{\lb}{\label}
\newcommand{\pr}{\prime}

\begin{document}
 \begin{center}
 {\Large\bf Path Integration for the Plane Pendulum with Finite Amplitude}
 \end{center}
 \vspace{5mm}
 \begin{center}
 H. Ahmedov$^1$  and I. H. Duru$^{2,1}$
 \end{center}
 1. Feza Gursey Institute,  P.O. Box 6, 81220,
 \c{C}engelk\"{o}y, Istanbul, Turkey \footnote{E--mail :
  hagi@gursey.gov.tr and duru@gursey.gov.tr}. \\
 2. Trakya University, Mathematics Department, P.O. Box
 126, Edirne, Turkey. \\

 \begin{center}
 {\bf Abstract}
 \end{center}
 Exact path  integration for the one dimensional potential $V=b^2\cos 2q$
which  describes the finite amplitude pendulum is presented.

 \vspace{5mm}
 \noindent
 {\bf 1. Introduction }

 \vspace{1mm}
 \noindent
 In last two decades almost all of the quantum mechanical
 potential problems which are solvable by the Schr$\ddot{o}$dinger
 equation method have also been solved by path integrals \cite{gr}.

 In this note we add the  finite amplitude plane pendulum
 potential $ V(q)= b^2\cos 2q$ to the list of the exact path
 integral solutions.

 \vspace{5mm}
 \noindent
 {\bf 2. Path Integral for the potential $V=b^2\cos 2q$ }

 \vspace{1mm}
 \noindent
 The Lagrangian path integral for the particle of the unit mass moving
 in the  potential
 \beq
 V= b^2\cos(2q), \ \ \ \ \  b^2\geq 0, \ \ \ q\in [0, 2\pi)
 \eeq
 is given by
 \beq\lb{1}
 K(q, q^\pr, T)=  \int  \mathcal{D}q \exp{(i \int_0^T
[\fr{1}{2}(\dot{q})^2- b^2\cos(2q)])}
 \eeq
 represents the probability amplitude for traveling  from a point $q$
 to $q^\pr$ in the time interval $T$. The explicit form of the
 expression (\ref{1}) is given by the usual time graded form as
 \cite{fey}
 \beq\lb{short}
 K(q, q^\pr, T)= \lim_{n\rightarrow \infty} (\fr{1}{2\pi i
\varepsilon})^{\fr{n+1}{2}}
 (\prod_{j=1}^n  \int_0^{2\pi}dq_j) \prod_{j=0}^n K_j
 \eeq
 where $q=q_0$, $q^\pr=q_{n+1}$, $\varepsilon=\fr{T}{n+1}$. $K_j$ is the
short time interval kernel
 between the space time points $(q_j, j \varepsilon)$ and $(q_j, (j+1)
 \varepsilon)$:
 \beq
 K_j= \exp{(\fr{i}{\varepsilon}[\fr{(q_j-q_{j+1})^2}{2}+(b\varepsilon)^2
 (2\sin q_j\sin q_{j+1}-1)])}.
 \eeq
 Writing the short time interval displacement ( in $\varepsilon\rightarrow
 0$ limit )as
 \beq
 \fr{(q_j-q_{j+1})^2}{2} \approx 1-\cos (q_j-q_{j+1})
 \eeq
 we get
 \beq\lb{kp}
 K_j= e^{\fr{i}{\varepsilon}(1-(b\varepsilon)^2)} P_j,
 \eeq
 where
 \beq
 P_j = \exp({-2ib [\cosh y\cos q_j\cos q_{j+1}+ \sinh y\sin q_j\sin
 q_{j+1}]}).
 \eeq
 The hyperbolic angle $y$ is defined ( in $\varepsilon\rightarrow 0$ limit
 ) as
 \beq
 \cosh y= \fr{1}{2b\varepsilon}, \ \ \ \sinh y \approx
 \fr{1}{2b\varepsilon}(1-2 (b\varepsilon)^2).
 \eeq
 Now we are ready to employ the well known  expansion formula
 \cite{erd}
 \begin{eqnarray}\lb{exp}
 P_j=2\sum_{m=0}^\infty [
 \fr{Ce_{2m}(y)}{p_{2m}} ce_{2m}(q_j)ce_{2m}(q_{j+1})+
 \fr{Se_{2m+2}(y)}{s_{2m+2}} se_{2m+2}(q_j)se_{2m+2}(q_{j+1})
 \nonumber \\
 -i\fr{Ce_{2m+1}(y)}{p_{2m+1}} ce_{2m+1}(q_j)ce_{2m+1}(q_{j+1})-i
 \fr{Se_{2m+1}(y)}{s_{2m+1}} se_{2m+1}(q_j)se_{2m+1}(q_{j+1})],
 \end{eqnarray}
 where $ce_n(x)\equiv ce_n(x, b^2)$ and $se_n(x)\equiv se_n(x, b^2))$
 are the even and odd periodic solutions of the Mathieu equation
 \beq
 (\fr{d^2}{dx^2}-2b^2)\cos 2x) y_n= -h_n y_n
 \eeq
 with $h_n$ being the corresponding eigenvalues. The coefficients
 in the expansion formula (\ref{exp}) are dependent on the eigenvalues of
the
 Mathieu differential equation [See Appendix].

 Inserting (\ref{kp}) and (\ref{exp}) into the full kernel (\ref{short})
we get
 \beq\lb{a}
 K(q, q^\pr, T)= \lim_{n\rightarrow \infty}
 (\fr{e^{\fr{i}{\varepsilon}(1-(b\varepsilon)^2)}}{\sqrt{2\pi i
\varepsilon}})^{n+1}
 (\prod_{j=1}^n \int_0^{2\pi}dq_j) \prod_{j=0}^n P_j.
 \eeq
 By making use of the  the orthogonality  relations
 \begin{eqnarray}
 \int_0^{2\pi} ce_n(x)ce_m(x) dx &=& \pi\delta_{nm}, \\
 \int_0^{2\pi} se_n(x)se_m(x) dx &=& \pi\delta_{nm}, \\
 \int_0^{2\pi} se_n(x)ce_m(x) dx &=& 0
 \end{eqnarray}
 we can execute the integration in (\ref{a}) and arrive at
 \beq
 K(q,q^\pr,T)=\sum_{m=0}^\infty  \Omega_{m}
\fr{ce_{m}(q)ce_{m}(q^\pr)}{\pi}+
 \sum_{m=1}^\infty \Xi_{m} \fr{se_{m}(q)se_{m}(q^\pr)}{\pi},
 \eeq
 where
 \begin{eqnarray}\lb{as}
 \Omega_{2m} &=& \lim_{n\rightarrow\infty} (
\sqrt{\fr{2\pi}{i\varepsilon}}
 \fr{Ce_{2m}(y)}{p_{2m}}
 e^{\fr{i}{\varepsilon}(1-(b\varepsilon)^2)})^{n+1},  \\
 \Omega_{2m+1} &=& \lim_{n\rightarrow\infty} (-i
\sqrt{\fr{2\pi}{i\varepsilon}}
 \fr{Ce_{2m+1}(y)}{p_{2m+1}}
 e^{\fr{i}{\varepsilon}(1-(b\varepsilon)^2)})^{n+1}  \\
 \Xi_{2m+2} &=& \lim_{n\rightarrow\infty} ( \sqrt{\fr{2\pi}{i\varepsilon}}
 \fr{Se_{2m+2}(y)}{s_{2m+2}}
 e^{\fr{i}{\varepsilon}(1-(b\varepsilon)^2)})^{n+1}  \\
 \Xi_{2m+1}&=& \lim_{n\rightarrow\infty} (-i
\sqrt{\fr{2\pi}{i\varepsilon}}
 \fr{Se_{2m+1}(y)}{s_{2m+1}}
 e^{\fr{i}{\varepsilon}(1-(b\varepsilon)^2)})^{n+1}.
 \end{eqnarray}
 Now we can calculate the  limits in the above coefficients:

 Using  the representation  for the  Mathieu function
 ( with $e^y= \fr{1}{b\varepsilon}$ in $\varepsilon\rightarrow 0$ limit  )
 \beq
 Ce_{2m}(y,  b^2)= \fr{p_{2m}}{A_0^{2m}}\sum_{r=0}^\infty (-)^r
 A^{2m}_{2r} J_r(be^{-y})J_r(be^y)
 \eeq
 and the asymptotic formula for the Bessel functions \cite{ed}
 \beq
 J_r(x)\approx i^r \fr{e^{-i(x+\fr{r^2-1/4}{2x}})}{\sqrt{-2\pi i x}}, \ \
\ x\rightarrow \infty
 \eeq
 and
 \beq
 J_r(x)\approx \fr{x^r}{2^r r!}, \ \ \ x\rightarrow 0
 \eeq
 we obtain
 \beq
 Ce_{2m}(y, b^2)\approx
 e^{(\fr{i\varepsilon}{8}-\fr{i}{\varepsilon})}
\sqrt{\fr{i\varepsilon}{2\pi}}
 \fr{p_{2m}}{A_0^{2m}}
 \sum_{r=0}^\infty (-i)^r A^{2m}_{2r} \fr{(b^2\varepsilon)^r}{2^r r!}
 e^{-i \fr{r^2}{2}\varepsilon}.
 \eeq
 Inserting the above formula into (\ref{as})  we have
 \beq
 \Omega_{2m}=e^{ih_0T}\lim_{n\rightarrow\infty} ( \omega_{2m}
 )^{n+1}.
 \eeq
 Here  $h_0=1/8-b^2$ and
 \beq
 \omega_{2m}= \fr{1}{A_0^{2m}}
 \sum_{r=0}^\infty (-i)^r A^{2m}_{2r} \fr{(b^2\varepsilon)^r}{2^r r!}
 e^{-i \fr{r^2}{2}\varepsilon}\approx
 1-i\fr{A_2^{2m}}{A_0^{2m}}\fr{b^2}{2} \varepsilon +
 O(\varepsilon^2)
 \eeq
 which from the recurrence relations (A.5) can be written as
 \beq\lb{z}
 \omega_{2m} \approx  1-i\fr{a_{2m}}{2} \varepsilon +
 O(\varepsilon^2).
 \eeq
 To obtain the energy spectrum we represent $\Omega_{2m}$ as
 \beq\lb{w}
 \Omega_{2m} = e^{i(h_0+E_{2m})T}= e^{ih_0T}\lim_{n\rightarrow\infty}
 (1+\fr{iTE_{2m}}{n+1})^{n+1}.
 \eeq
 Comparing (\ref{w}) to (24) ( with $T=(n+1)\varepsilon$ ) and using
(\ref{z}) we
 get the energy spectrum for the wave functions $ce_{2m}(q)$
 \beq
 \mathcal{E}_{2m}=h_0+E_{2m}=h_0-\fr{a_{2m}}{2}.
 \eeq
 In the  similar fashions using  the representations \cite{erd}
 \beq
 Ce_{2m+1}(y)=\fr{p_{2m+1}}{A^{2m+1}_1}\sum_{r=0}^\infty (-)^r
 A^{2m+1}_{2r+1} [J_r(be^{-y})J_{r+1}(be^y) + J_r(be^{y})J_{r+1}(be^{-y})]
 \eeq
 \beq
 Se_{2m+1}(y)=-\fr{s_{2m+1}}{iB^{2m+1}_1}\sum_{r=0}^\infty (-)^r
 B^{2m+1}_{2r+1} [J_r(be^{-y})J_{r+1}(be^y) - J_r(be^{y})J_{r+1}(be^{-y})]
 \eeq
 \beq
 Se_{2m+2}(y)=\fr{s_{2m+2}}{iB^{2m+2}_2}\sum_{r=0}^\infty (-)^r
 B^{2m+2}_{2r+2} [J_r(be^{-y})J_{r+2}(be^y) - J_r(be^{y})J_{r+2}(be^{-y})]
 \eeq
 we obtain the remaining coefficients
 \begin{eqnarray}
 \Omega_{2m+1} &=& e^{i(h_0-a_{2m+1}/2)T}, \\
 \Xi_{2m+1} &=& e^{i(h_0-b_{2m+1}/2)T}, \\
 \Xi_{2m+2} &=& e^{i(h_0-b_{2m+2}/2)T}
 \end{eqnarray}
 which exhibit the full energy spectrum.

 The coordinate $q$ and time $t$ in the above derivation are
dimensionless.
 To introduce dimension we have to make the following replacements
 \beq
 t\rightarrow \mu t, \ \ \ q\rightarrow \mu q,
 \eeq
 where $\mu$ is the mass of the pendulum.

 When the sign of $b^2$ reversed to $b^2<0$ one only needs to
 replace the argument $q$ of the Mathieu functions by
 $q=\fr{\pi}{2}+q$, the energy spectrum remains unchanged.

 \setcounter{equation}{0} \def\theequation{A.\arabic{equation}}
 \begin{center}
 {\bf Appendix \cite{erd} }
 \end{center}
 $Ce_n(x)$ and $Se_n(x)$ in the formula  (\ref{exp}) are called associated
Mathieu functions
 and defined as $Ce_n(x)=ce_n(ix)$ and $Se_n(x)=se_n(ix)$. The constants
$p_n$
 and $s_n$ are defined as
 \beq
 p_{2m}= \fr{ce_{2m}(0)ce_{2m}^\pr(\pi/2)}{A_0^{2m}},
 \eeq
 \beq
 p_{2m+1}= -\fr{ce_{2m}(0)ce_{2m}^\pr(\pi/2)}{bA_1^{2m+1}},
 \eeq
 \beq
 s_{2m+1}= \fr{se_{2m+1}^\pr(0)se_{2m+1}(\pi/2)}{b B_1^{2m+1}},
 \eeq
 \beq
 s_{2m+2}= \fr{se_{2m+2}^\pr(0)se_{2m+2}(\pi/2)}{b^2 B_2^{2m+2}},
 \eeq
 where $A^n_r$ and $B^n_r$ satisfy the following recurrence
 relations
 \begin{eqnarray}
  a_{2m}A^{2m}_0-b^2A_{2}^{2m}=0 \ \ \  \ \ \ \ \ \ \ \ \ \ \ \ \ \ \ \ \
\ \ \nonumber \\
 (a_{2m}-4r^2) A^{2m}_{2r}-b^2(A_{2r+2}^{2m}+A_{2r-2}^{2m})=0, \ \
 \ r\geq 1,
 \end{eqnarray}
 \begin{eqnarray}
 (a_{2m+1}-1-b^2)A^{2m+1}_1-b^2A_{3}^{2m+1}=0 \ \ \  \ \ \ \ \ \ \ \ \ \ \
\ \ \ \ \ \ \ \ \nonumber \\
 (a_{2m+1}-(2r+1)^2)
A^{2m+1}_{2r+1}-b^2(A_{2r+3}^{2m+1}+A_{2r-1}^{2m+1})=0, \ \
 \ r\geq 1
 \end{eqnarray}
 \begin{eqnarray}
 (b_{2m+1}-1-b^2)B^{2m+1}_1-b^2B_{3}^{2m+1}=0 \ \ \  \ \ \ \ \ \ \ \ \ \ \
\ \ \ \ \ \ \ \ \nonumber \\
 (b_{2m+1}-(2r+1)^2)
B^{2m+1}_{2r+1}-b^2(B_{2r+3}^{2m+1}+B_{2r-1}^{2m+1})=0, \ \
 \ r\geq 1
 \end{eqnarray}
 \begin{eqnarray}
 (b_{2m}-4)B^{2m+1}_2-b^2B_{4}^{2m+2}=0 \ \ \  \ \ \ \ \ \ \ \ \ \ \ \ \ \
\ \ \ \ \ \nonumber \\
 (b_{2m}-4r^2) B^{2m+2}_{2r}-b^2(B_{2r+2}^{2m+2}-B_{2r-2}^{2m+2})=0, \ \
 \ r\geq 2
 \end{eqnarray}
 and normalization
 \beq
 \sum_{r=0}^\infty A^{2m}_{2r} > 0, \ \ \ \
 2(A^{2m}_0)^2+\sum_{r=0}^\infty (A^{2m}_{2r})^2=1
 \eeq
 \beq
 \sum_{r=0}^\infty A^{2m+1}_{2r+1} > 0, \ \ \ \
 \sum_{r=0}^\infty (A^{2m+1}_{2r+1})^2=1
 \eeq
 \beq
 \sum_{r=0}^\infty (2r+1)B^{2m+1}_{2r+1} > 0, \ \ \ \
 \sum_{r=0}^\infty (B^{2m+1}_{2r+1})^2=1
 \eeq
 \beq
 \sum_{r=0}^\infty (2r+2)B^{2m+2}_{2r+2} > 0, \ \ \ \
 \sum_{r=0}^\infty (B^{2m+2}_{2r+2})^2=1.
 \eeq
 Here $a_{2n}$, $a_{2n+1}$, $b_{2n}$ and $b_{2n+1}$ are the  eigenvalues
of $ce_{2m}$, $ce_{2n+1}$, $se_{2n}$ and
 $se_{2n+1}$.

 \vspace{4mm}
 \noindent
 Acknowledgment: We thank Simon Bolivar University ( Caracas )
 where this work was initiated while one of the authors ( I.H.
 Duru ) was visiting there.

 \end{document}